\documentclass[preprint]{revtex4}
\usepackage{epsfig}

\begin{document}

\title{Phase diagram of a Bose gas near a wide Feshbach resonance}
\author{Lan Yin}
\email{yinlan@pku.edu.cn} \affiliation{School of Physics, Peking
University, Beijing 100871, China}

\begin{abstract}
In this paper, we study the phase diagram of a homogeneous Bose gas
with a repulsive interaction near a wide Feshbach resonance at zero
temperature.  The Bose-Einstein-condensation (BEC) state of atoms is
a metastable state.  When the scattering length $a$ exceeds a
critical value depending on the atom density $n$, $na^3>0.035$, the
molecular excitation energy is imaginary and the atomic BEC state is
dynamically unstable against molecule formation.  The BEC state of
diatomic molecules has lower energy, where the atomic excitation is
gapped and the molecular excitation is gapless. However when the
scattering length is above another critical value, $na^3>0.0164$,
the molecular BEC state becomes a unstable coherent mixture of atoms
and molecules.  In both BEC states, the binding energy of diatomic
molecules is reduced due to the many-body effect.
\end{abstract}

\maketitle
\section{Introduction}

Bose gases with Feshbach resonances have shown very interesting
properties.  The particle loss rate due to the three-body
recombination increases enormously near a Feshbach resonance
\cite{Stenger}. A sudden change in the magnetic field generates
oscillations between atoms and diatomic molecules \cite{Donley}.
Diatomic molecules can be produced when the magnetic field is either
tuned through the resonance \cite{Xu} or oscillates with a frequency
corresponding to the molecular binding energy \cite{Thompson}.
Recently observed Effimov effect \cite{Kraemer} may help to control
the particle loss for more extensive studies near a Feshbach
resonance.

The atomic BEC state with strong interactions was explored in a
variational approach \cite{Cowell}.  At resonance the energy per
atom was found to be proportional to $n^{2/3}$ \cite{Cowell}.  A
transition between atomic and molecular BEC states at the resonance
was proposed \cite{Radzihovsky, Romans}. However, these states were
found unstable in some regions due to negative compressibility
\cite{Basu}. Recently, a transition between the molecular BEC state
and a coherent mixture state of atoms and molecules was proposed
\cite{Braaten1}.

In a Bose gas with a Feshbach resonance, scattering states in the
open channel are coupled to bound states in the closed channel.
Eigenstates of diatomic molecules are superpositions of both open
and closed channel states. Near a wide resonance where the effective
range of the interaction is very small, the magnitude of the
close-channel component is much less than that of the open-channel
component, which can be seen in the renormalization of the molecular
propagator \cite{Yin}.   Thus an effective model of open-channel
atoms should be able to describe the system with a wide resonance. A
single-channel model has been widely used to describe both normal
and BEC states of Bose gases \cite{Fetter}, given by
\begin{equation}\label{H}
H=-\frac{\hbar^2}{2m}\psi^{\dagger}\nabla^2\psi+{g \over
2}\psi^{\dagger}\psi^{\dagger}\psi \psi,
\end{equation}
where $\psi$ is atomic field operator, $g$ is the coupling constant,
$g \equiv 4\pi\hbar^2a/m$, and $a$ is the scattering length.  When
the scattering length is negative, $a<0$, the system is unstable and
subject to mechanical collapse at low temperatures \cite{Stoof,
Mueller, Jeon}.  Therefore in the following we focus on the region
with repulsive interactions, $a>0$.

A crucial question about the single-channel model is whether or not
it is capable of describing diatomic molecules near the resonance.
The answer of this question is positive, because the Hamiltonian
given by Eq. (\ref{H}) has bound eigenstates of two atoms at $a>0$.
These bound states are the eigenstates of diatomic molecules, given
by
\begin{equation}
|\phi_{\bf p}\rangle={1\over \sqrt{V}}\sum_{\bf q}{{\cal N} \over
E_a+2\epsilon_q}\psi^\dagger_{{\bf q+p}/2}\psi^\dagger_{{\bf
-q+p}/2}|0\rangle,
\end{equation}
where $V$ is the volume, ${\cal N}$ is a normalization constant,
$\epsilon_q\equiv\hbar^2q^2/(2m)$, and the molecular binding energy
in vacuum is given by $E_a=\hbar^2/(ma^2)$.  The energy eigenvalue
of the molecule is given by $-E_a+\epsilon_p/2$, consistent with the
energy of shallow bound states in the quantum scattering theory
\cite{Sakurai}. It is important to note that the single-channel
model is an effective field-theory model and ultraviolet divergences
in calculations need to be removed by the correct renormalization
scheme \cite{Braaten}.

\begin{figure}\centering
\includegraphics[width=3in]{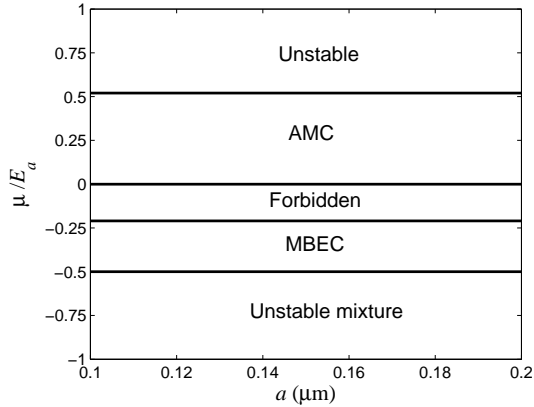}
\caption{The phase diagram of a Bose gas near a wide Feshbach
resonance at zero temperature.  The atomic BEC state (ABEC) is
metastable with $0<\mu<0.52E_a$.  The molecular BEC (MBEC) state is
with $-0.5E_a<\mu<-0.21E_a$. }\label{Pd}
\end{figure}

In this paper, the phase diagram of a homogeneous Bose gas near a
wide Feshbach resonance is obtained at zero temperature, as shown in
Fig. \ref{Pd}. The atomic BEC state is a metastable state when the
chemical potential $\mu$ is between $0$ and $0.52 E_a$,
corresponding to $0<na^3<0.035$. The atomic BEC state is unstable
against molecule formation when $na^3>0.035$. The molecular BEC
state is stable when $-0.21E_a>\mu>-0.5E_a$, or $0.0164>na^3>0$.
There is no solution when $-0.21 E_a<\mu<0$. The coherent mixture
state of atoms and molecules is subject to mechanical collapse,
although $\mu$ can be smaller than $-0.5 E_a$ at very high
densities.  In both BEC states, the molecular binding energy is
smaller than $E_a$ due to interaction. This phase diagram is valid
when $a$ is much larger than the effective range.

\section{The atomic BEC state}
Bose atoms condense below the BEC transition temperature.  The
atomic BEC state in the dilute case is well described by
Bogoliubov's theory in which the atomic field operator has a finite
expectation value, $\langle\psi\rangle=\psi_0$.  In the grand
canonical ensemble, the grand thermodynamic potential $F \equiv
H-\mu \psi^\dagger\psi$ has a constant part $F_0=gn_0^2/2 -\mu n_0$,
where $\mu$ is the chemical potential and $n_0\equiv|\psi_0|^2$. The
saddle-point condition $\delta F_0/\delta\psi_0$ yields $\mu=g n_0$.
For convenience $\psi_0$ can be chosen to be positive.  The
quadratic part of the grand potential describes the gaussian
fluctuation in the atomic field $\delta\psi \equiv \psi-\psi_0$,
\begin{equation}
F_2=-\frac{\hbar^2}{2m}\delta\psi^{\dagger}\nabla^2\delta\psi+
gn_0\delta\psi^\dagger\delta\psi+ {1 \over 2}
gn_0(\delta\psi^\dagger\delta\psi^\dagger+h.c.).
\end{equation}
By Bogoliubov transformation, the quadratic grand potential $F_2$
can be diagonalized,
\begin{equation}
F_2=C+\sum_{\bf k}E_k c_{\bf k}^\dagger c_{\bf k},
\end{equation}
where the quasi-particle energy is given by $E_k=\sqrt{\epsilon_k
(\epsilon_k+2gn_0)}$, $\epsilon_k \equiv \hbar^2k^2/(2m)$, the
quasi-particle operator is given by $c_{\bf k} \equiv u_k \psi_{\bf
k}-v_k \psi_{-{\bf k}}^\dagger$,
$u_k^2=[1+(\epsilon_k+gn_0)/E_k]/2$, and $v_k^2=u_k^2-1$.  The
quantum depletion, the vacuum of quasi-particles, contributes to the
ground-state energy by $C=8gn_0(m gn_0/\hbar^2)^{3/2}/(15\pi^2)$.
The higher order terms in the grand potential, $$
g\delta\psi^\dagger\delta\psi[(\psi_0 \delta\psi+h.c) +
\delta\psi^\dagger\delta\psi/2],$$ are neglected in Bogoliubov's
approximation.  In Popov's approximation \cite{Popov}, the
mean-field terms of non-condensed atoms are added in the
particle-hole channel, and the chemical potential is shifted,
$\mu=g(n_0+2\delta n)$, where $\delta n=8n_0\sqrt{n_0a^3/\pi}/3$ is
the the atom density of the quantum depletion. In the diagrammatic
approach in the traditional theory of a dilute Bose gas
\cite{Fetter}, thermodynamical properties can be calculated by
perturbation in the order of $\sqrt{na^3}$ beyond Bogoliubov's
approximation.

The traditional theory is accurate in the dilute case, where the
perturbation can be stopped at a sufficient order of $\sqrt{na^3}$
without losing much accuracy.  However near the resonance, there are
strong fluctuations around the condensate.  In addition to
single-atom excitations, collective excitations must be considered,
such as excitations of diatomic molecules.  The dispersion of
molecular excitations can be obtained from the poles of the
two-particle correlation function of non-condensed atoms.  In the
atomic BEC state, the particle-particle channel and the
particle-hole channel are coupled, and the correlation function is a
$3\times3$ matrix given by
\begin{equation}
\chi_{\alpha\beta}({\bf r}-{\bf r}',t-t') \equiv -{i \over \hbar}
\langle T[ b_\alpha({\bf r},t) b_\beta^\dagger({\bf r}',t')]
\rangle,
\end{equation}
where $b_1=\psi^2$, $b_2=b_1^\dagger$, $b_3=2 \psi^\dagger \psi$,
and $T$ is the time-ordering operator.

Due to the strong interaction near the resonance, any perturbation
theory truncated at any finite order of $\sqrt{na^3}$ will fail.  In
the following, we adopt the Random Phase Approximation (RPA) which
focuses on the renormalization to the two-body interaction.  In RPA,
there are Feynman diagrams from all orders of perturbation and the
result can often be applied to the strong-interaction region with
correct qualitative features \cite{Fetter}.

In RPA, the correlation function is given by
\begin{equation}\label{cdk}
\chi({\bf k},\omega)=[1-g \chi^{(0)}({\bf k},\omega)]^{-1}
\chi^{(0)}({\bf k},\omega).
\end{equation}
The function $\chi^{(0)}({\bf k},\omega)$ is the correlation
function calculated in Bogoliubov's approximation, given by
\begin{eqnarray}
\chi_{11}^{(0)}({\bf k},\omega)&=&\int {d^3k' \over (2\pi)^3}
({u_{k'}^2 u_{|{\bf k}-{\bf k'}|}^2 \over A({\bf k},{\bf
k'},\omega)}+{v_{k'}^2 v_{|{\bf k}-{\bf k'}|}^2
\over A({\bf k},{\bf k'},-\omega)}),\nonumber\\
\chi_{12}^{(0)}({\bf k},\omega)&=&2\int {d^3k' \over (2\pi)^3}
{u_{k'} v_{k'} u_{|{\bf k}-{\bf k'}|} v_{|{\bf k}-{\bf k'}|}
\over B({\bf k},{\bf k'},\omega)}, \nonumber \\
\chi_{13}^{(0)}({\bf k},\omega)&=&2\int {d^3k' \over (2\pi)^3}({
u_{k'} v_{k'} u^2_{|{\bf k}-{\bf k'}|} \over A({\bf k},{\bf
k'},\omega)} +{u_{k'} v_{k'} v^2_{|{\bf k}-{\bf k'}|} \over
A({\bf k},{\bf k'},-\omega)}), \nonumber \\
\chi_{33}^{(0)}({\bf k},\omega)&=&2\int {d^3k' \over (2\pi)^3}
{(u_{k'}v_{|{\bf k}-{\bf k'}|}+ v_{k'}u_{|{\bf k}-{\bf k'}|})^2
\over B({\bf k},{\bf k'},\omega)},\label{chi0}
\end{eqnarray}
$\chi_{22}^{(0)}({\bf k},\omega)=\chi_{11}^{(0)}(-{\bf k},-\omega)$,
$\chi_{23}^{(0)}({\bf k},\omega)=\chi_{13}^{(0)}(-{\bf k},-\omega)$,
and $\chi_{ij}^{(0)}({\bf k},\omega)=\chi_{ji}^{(0)}({\bf
k},\omega)$ for $j\neq i$, where $A({\bf k},{\bf k'},\omega)\equiv
\hbar\omega-E_k-E_{|{\bf k}-{\bf k'}|}+i\delta$ and $1/B({\bf
k},{\bf k'},\omega)\equiv 1/A({\bf k},{\bf k'},\omega)+1/A({\bf
k},{\bf k'},-\omega)$.

\begin{figure}\centering
\includegraphics[width=3in]{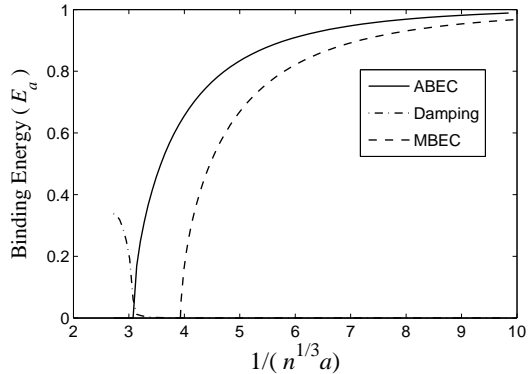}
\caption{The binding energy and damping of diatomic molecules versus
$n^{1/3}a$ in atomic and molecular BEC states. The solid line is the
binding energy and the dash-dotted line is the imaginary or damping
part of the molecular excitation in the atomic BEC state.  The
dashed line is the binding energy in the molecular BEC
state.}\label{Eng}
\end{figure}

The dispersion of diatomic molecules can be obtained from the pole
of the correlation function given in Eq. (\ref{cdk}),
\begin{equation}\label{pole}
\det|{\rm I}-g\chi^{(0)}({\bf k},\omega)|=0,
\end{equation}
where ${\rm I}$ is the identity matrix.  The real part of the
molecular excitation energy at ${\bf k}=0$ is the negative of the
binding energy which is plotted in Fig. \ref{Eng}. When the
interaction is weak, $na^3\ll1$, the molecular binding energy is
approximately given by $E_a-2gn_0$, which recovers the vacuum result
in the dilute limit.  The size of the bound state is of the order of
the scattering length $a$.  When the wavevector $k$ is of the order
of $1/a$ or larger, the atomic-excitation energy is approximately
given by $\epsilon_k+gn_0$.  This energy shift $gn_0$ in atomic
excitation energy at high momentum causes the reduction in the
molecular binding energy by $-2gn_0$.

When the interaction is stronger, the imaginary part of the
molecular excitation energy becomes noneligible.  At the critical
value of $na^3= 0.035$ or $\mu=0.52E_a$, the real part of the
molecular energy vanishes and only the imaginary part is finite,
implying that beyond this point the system is totally unstable
against molecule formation.  This instability is a many-body effect,
due to the process described in RPA in which two atoms from two
pairs in the quantum depletion form a molecule leaving the rest two
atoms in the two pairs excited. Such process is absent in the normal
state where there is no off-diagonal correlation.  In contrast, the
particle loss due to the three-body recombination is a few-body
effect which is present in both BEC and normal states.

\section{The molecular BEC state}

When the scattering length is positive, $a>0$, diatomic molecules
have lower energy than atoms, and can condense into a molecular BEC
state at low temperatures. In the single-channel model, the
molecular condensation can be described by the off-diagonal long
range order $\Delta \equiv g \langle \psi\psi \rangle \neq 0$. For
simplicity, we assume $\Delta>0$.   The mean-field grand potential
describing the molecular BEC state is given by
\begin{equation}
F_m=F_m^{(0)}-\frac{\hbar^2}{2m}\psi^{\dagger}\nabla^2\psi+(2gn-\mu)\psi^{\dagger}\psi+
{\Delta\over 2}(\psi^\dagger\psi^\dagger+\psi\psi),
\end{equation}
where $F_m^{(0)}=-(gn^2+\Delta^2/2g)$. This grand potential $F_m$
can be diagonalized by Bogoliubov transformation,
\begin{equation}
F_m=F_{m0}+\sum_{\bf k}E_k c_{\bf k}^\dagger c_{\bf k},
\end{equation}
where $E_k=\sqrt{(\epsilon_k+2gn-\mu)^2-\Delta^2}$, $c_{\bf k}=u_k
\psi_{\bf k}-v_k \psi_{-{\bf k}}^\dagger$,
$u_k^2=[1+(\epsilon_k+2gn-\mu)/E_k]/2$, $v_k^2=u_k^2-1$, and the
constant $F_{m0}$ is the ground state energy£¬$$ F_{m0}=\sum_{\bf
k}{1\over 2}[E_k-(\epsilon_k+2gn-\mu)+{\Delta^2 \over 2E_k}]-gn.$$
The parameter $\Delta$ can be determined self-consistently,
$$\Delta=g\int u_kv_k {d^3k \over(2\pi)^3},$$ or
\begin{equation} {1\over g}=\int {d^3k\over(2\pi)^3}[-{1\over
2E_k}+{1\over 2\epsilon_k}],  \label{gap}
\end{equation}
where the last term in the integrand $1/(2\epsilon_k)$ is a counter
term. The chemical potential $\mu$ and parameter $\Delta$ can be
solved from Eq. (\ref{gap}) and the equation for the density $n$,
\begin{equation}
n=\int{d^3k\over(2\pi)^3}v_k^2=\int{d^3k\over(2\pi)^3}[{\epsilon_k+2gn-\mu
\over 2E_k}-1]. \label{n}
\end{equation}

There is a gap in the atomic excitation given by
$E_0=\sqrt{(2gn-\mu)^2-\Delta^2}$, consistent with the fact that
atoms have higher energy than molecules.  The molecular binding
energy in the molecular BEC state is given by $2E_0$, equal to the
energy difference between two atomic excitations and a molecular
excitation at ${\bf k}=0$.  The binding energy is plotted in Fig.
\ref{Eng}. In the dilute limit where $na^3\ll1$, the gap $E_0$ is
approximately equal to $ E_a/2$. However at the critical value of
$na^3 = 0.0164$, the gap vanishes. When $na^3
> 0.0164$, there is no solution, indicating that the molecular BEC
state no longer exists.

Similar to the case in the atomic BEC state, the dispersion of
molecular excitations in the molecular BEC state can be obtained
from the poles of the two-particle correlation function $\chi$.  In
RPA, the pole is given by Eq. (\ref{pole}), and the mean-field
correlation function $\chi^{(0)}$ is given by Eq. (\ref{chi0}), with
the coefficients, $u_k$ and $v_k$, and the excitation energy $E_k$
replaced with the values in the molecular BEC state. When ${\bf
k}=0$ and $\omega=0$, Eq. (\ref{pole}) is automatically satisfied
following Eq. (\ref{chi0}, \ref{gap}), showing that the molecular
excitation is gapless. At small $k$ and $\omega$, to the leading
order of $k$, the molecular excitation frequency is linearly
dispersed, $\omega_k \approx v_m k$, where $v_m$ is the molecule
velocity.

\begin{figure}\centering
\includegraphics[width=3in]{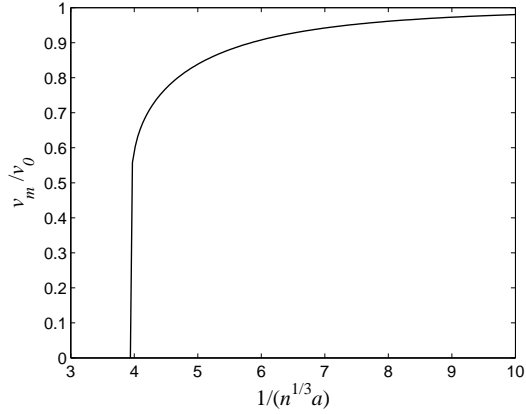}
\caption{The molecular velocity $v_m$ versus $n^{1/3}a$ in the
molecular BEC state.  At $na^3 = 0.0164$, the velocity $v_m$
vanishes.}\label{vm}
\end{figure}

The molecule velocity is plotted in Fig. \ref{vm}.  In the dilute
limit when $na^3\ll1$, the molecule velocity is approximately given
by
\begin{equation} \label{v0}
v_0={\hbar\over m} \sqrt{3\pi na}.
\end{equation}
If Eq. (\ref{v0}) is compared to the phonon velocity $v_p=\hbar
\sqrt{4\pi na}/m$ in the atomic BEC state, the naive estimation of
the molecule-molecule scattering length is $6a$, larger than the
result from solving the four-body problem \cite{Petrov}.  This
reason for this discrepancy might be that in the dilute limit the
mean-field density given by Eq. (\ref{n}) is smaller than the true
density due to fluctuations, similar to that in the molecular BEC
state of the Fermi gas \cite{NSR}. When the interaction is stronger,
at the critical value of $na^3 = 0.0164$ or $\mu=-0.21E_a$, the
molecule velocity $v_m$ drops to zero, which is very similar to
softening of phonon modes in solids near structural phase
transitions.  The vanish of both the atomic excitation gap $E_0$ and
molecule velocity $v_m$ at $na^3 = 0.0164$ indicates that beyond
this critical point the molecular BEC state does no exist.

\section{The coherent mixture of atoms and molecules}
In both BEC states, the molecular binding energy is reduced by the
many-body interaction. In the region beyond these BEC states, the
system may contain both atoms and molecules. Here we consider a
coherent mixture state of the atomic condensation with $\psi_0
\equiv \langle\psi\rangle$, and molecular condensation with $\Delta
\equiv g \langle \delta\psi\delta\psi \rangle$, where
$\delta\psi\equiv \psi-\psi_0$. In this mixture state, the two
parameters $\psi_0$ and $\Delta$ are in principle independent,
whereas in the atomic BEC state the parameter $\Delta$ is a function
of $\psi_0$.

In the mean-field approximation, the grand potential is given by
\begin{eqnarray}
F_{mx}=&&F_{mx}^{(0)}-{\hbar^2 \over
2m}\delta\psi^{\dagger}\nabla^2\delta\psi
+(2gn-\mu)\delta\psi^{\dagger}\delta\psi \nonumber \\
&&+{1\over 2}[(\Delta+g \psi_0^2)
\delta\psi^\dagger\delta\psi^\dagger+h.c],\label{mx}
\end{eqnarray}
where $F_{mx}^{(0)}=-[g\delta n^2+n_0(\mu-gn_0/2)+|\Delta|^2/
(2g)]$, $n_0\equiv |\psi_0|^2$, $\delta n \equiv \langle
\delta\psi^\dagger\delta\psi\rangle$, and $n=n_0+\delta n$. Its
mean-field expectation value, $\bar{F}_{mx}\equiv\langle
F_{mx}\rangle$, is given by
\begin{equation}
\bar{F}_{mx}={g\over2}|\psi_0^2+{\Delta \over g}|^2+2gn_0\delta
n+{g\over2}\delta n^2-\mu(\delta n+n_0).
\end{equation}
The parameter $\psi_0$ should minimize the grand potential,
$\partial\bar{F}_{mx}/\partial\psi_0^*=0$, which yields the
saddle-point equation
\begin{equation} \label{mumx}
\mu=g(n_0+2\delta n)+\Delta {\psi_0^* \over \psi_0}.
\end{equation}
Apparently, the product $\Delta {\psi_0^*}^2$ is real. For
simplicity, we choose $\Delta$ to be positive, which also means that
$\psi_0^2$ is real.

The mean-field grand potential can be diagonalized by Bogoliubov
transformation, where the field operator of the quasi-particles are
given by $c_{\bf k}=u_k \psi_{\bf k}-v_k \psi_{-{\bf k}}^\dagger$,
the coefficients are given by $u_k^2=[1+(\epsilon_k+2gn-\mu)/E_k]/2$
and $v_k^2=u_k^2-1$, and the quasi-particle energy is given by
$E_k=\sqrt{(\epsilon_k+2gn-\mu)^2-(\Delta+g\psi_0^2)^2}$.  The
parameter $\Delta$ can be determined from the self-consistency
equation, $$\Delta=g\int{d^3k\over(2\pi)^3}u_kv_k,$$ i. e.
\begin{equation}
{1\over g}-{\psi_0^2 \over \Delta+g\psi_0^2}=\int
{d^3k\over(2\pi)^3}[-{1\over 2E_k}+{1\over 2\epsilon_k}].
\label{gapp}
\end{equation}
The expression of the total density is now given by
\begin{equation}\label{nmx}
n=n_0+\int {d^3k \over (2\pi)^3}{1\over2}({\epsilon_k+2gn-\mu \over
E_k}-1).
\end{equation}

\begin{figure}\centering
\includegraphics[width=3in]{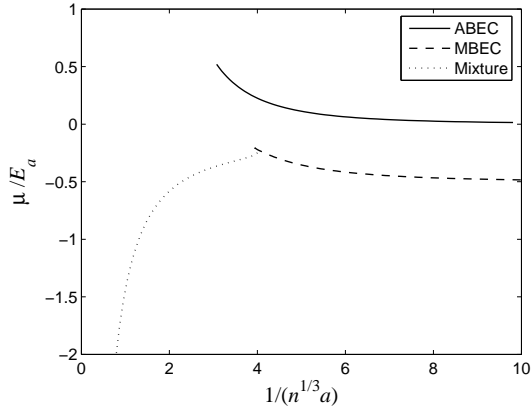}
\caption{The chemical potential $\mu$ versus $n^{1/3}a$. The solid
line is $\mu$ in the atomic BEC state in Popov's approximation
\cite{Popov} for $na^3<0.035$. The dashed line is $\mu$ in the
molecular BEC state which exists at $na^3<0.0164$.  The dotted line
is $\mu$ in the mixture state.}\label{mu}
\end{figure}

For a fixed total density $n$, both $\Delta$ and $\psi_0$ can be
solved from Eq. (\ref{mumx}, \ref{gapp}, \ref{nmx}). In the
solution, $\Delta$ and $\psi_0^2$ have opposite signs, whereas in
the atomic BEC state $\Delta$ and $\psi_0^2$ have the same sign.  At
the transition point to the molecular BEC state, the atomic
condensation density $n_0$ is zero.  The solution shows a reentrant
behavior in a tiny region with $0.0156<na^3<0.0164$. When
$na^3>0.0164$, the solution is unique.  However, in this region as
shown in Fig. \ref{mu}, the chemical potential $\mu$ decreases with
the increase in density $n$, which means that the compressibility
$\partial \mu/\partial n$ is negative and the mixture state is
subject to mechanical collapse. This instability is very similar to
the instability of Bose gases with attractive interactions
\cite{Stoof, Mueller, Jeon}. In a trap, the mixture state may be
stabilized by the finite-size effect under certain conditions, as
discovered in trapped Bose gases with attractive
interactions\cite{Ruprecht}.

{\bf Conclusion and Discussions}.  The phase diagram of a
homogeneous Bose gas near a wide Feshbach resonance is studied at
zero temperature.  In the atomic BEC state, the real part of the
molecular excitation energy vanish at $na^3=0.035$, implying the
instability against molecular formation for stronger interactions.
In the molecular BEC state, the atomic excitation energy is gapped
and the molecular excitation energy is linearly dispersed.  Both the
gap and molecular velocity vanish at $na^3=0.0164$, above which the
molecular BEC state no longer exists. In both BEC states molecular
binding energies are reduced by the many-body interaction.  In the
coherent mixture state of atoms and molecules, at $na^3>0.0164$, the
compressibility is found to be negative, indicating that the mixture
is subject to mechanical collapse, similar to Bose gases with
attractive interactions.

It is an open question whether or not there are more exotic states
in regions beyond the BEC states. So far three-body interactions are
ignored in the BEC states.  Although they are crucial to the
dynamical properties such as the particle-loss rate and the
stability time, the quasi-equilibrium properties is unlikely
affected. However triatomic Effimov molecules may form near the
resonance due to three-body interactions.  Whether or not a gas of
Effimov triatomic molecules can exist is at present unknown.

{\bf Acknowledgement}. We would like to thank D. J. Thouless, T.-L.
Ho, and H. Zhai for helpful discussions. This work is supported by
NSFC under Grant No. 90303008 and 10674007, and by Chinese MOST
under grant number 2006CB921401.


\begin{thebibliography}{99}
\bibitem{Stenger}J. Stenger \textit{et al}., Phys. Rev. Lett. \textbf{82}, 2422 (1999).
\bibitem{Donley} E. A. Donley \textit{et al}., Nature \textbf{417}, 529 (2002).
\bibitem{Xu} K. Xu \textit{et al}., Phys. Rev. Lett. \textbf{91}, 210402 (2003).
\bibitem{Thompson} S. T. Thompson, E. Hodby, and C. E. Wieman, Phys. Rev. Lett. \textbf{95}, 190404 (2005).
\bibitem{Kraemer} T. Kraemer \textit{et al}., Nature \textbf{440}, 315 (2006).
\bibitem{Cowell} S. Cowell \textit{et al}., Phys. Rev. Lett. \textbf{88}, 210403
(2002).
\bibitem{Radzihovsky} L. Radzihovsky, J. Park, and P. B. Weichman,
Phys. Rev. Lett. \textbf{92}, 160402 (2004).
\bibitem{Romans} M. W. J. Romans, R. A. Duine, S. Sachdev, and H. T. C. Stoof, Phys. Rev. Lett. {\bf
93}, 020405 (2004).
\bibitem{Basu} S. Basu and E. J. Mueller, cond-mat/0507460 (2005).
\bibitem{Braaten1} E. Braaten and D. Zhang, cond-mat/0703308 (2007).
\bibitem{Yin} L. Yin and Z.-H. Ning, Phys. Rev. A 68, 033608
(2003); Y. Zhang and L. Yin, Phys. Rev. A 72, 043607 (2005).
\bibitem{Popov} V. N. Popov, {\it Functional integrals
and collective excitations} (Cambridge University, Cambridge, 1987).
\bibitem{Fetter} For example, A. L. Fetter and J. D. Walecka, {\it Quantum theory of many-particle systems}
(McGraw-Hill, San Francisco, 1971); D. Pines, {\it The many-body
problem; a lecture note and reprint volume} (New York, W.A.
Benjamin, 1961).
\bibitem{Stoof} H. T. C. Stoof, Phys. Rev. A {\bf 49}, 3824 (1994).
\bibitem{Mueller} E. J. Mueller and G. Baym, Phys. Rev. A {\bf 62}, 053605 (2000).
\bibitem{Jeon} G. S. Jeon, L. Yin, S. W. Rhee, and D. J. Thouless, Phys. Rev. A {\bf 66}, 011603(R) (2002).
\bibitem{Sakurai}  J. J. Sakurai and S. F. Tuan, {\it Modern quantum mechanics} (Reading, Addison-Wesley,
1994.), chaptor 7.
\bibitem{Braaten} E. Braaten and A. Nieto, Phys. Rev. B 55, 8090
(1997).
\bibitem{Petrov} D. S. Petrov, C. Salomon, and G. V. Shlyapnikov, Phys. Rev. Lett. \textbf{93}, 090404 (2004).
\bibitem{NSR} P. Nozi$\grave{\rm e}$res and S. Schmitt-Rink, J. Low Temp. Phys. 59, 195 (1985); C. A. R. S$\acute{\rm a}$ de Melo, M. Randeria, and J. R.
Engelbrecht, Phys. Rev. Lett. \textbf{71}, 3202 (1993).
\bibitem{Ruprecht} P. A. Ruprecht, M. J. Holland, K. Burnett, and M. Edwards, Phys. Rev. A {\bf 51}, 4704 (1995).
\end{thebibliography}
\end{document}